\documentclass[a4paper,12pt]{article}
\usepackage{amsmath,amssymb,mathtools} 
\usepackage{color}
\usepackage{graphicx}
\usepackage{subfigure}
\usepackage{cite}           
\usepackage{hyperref}            
\usepackage{multirow,makecell}   
\usepackage{textcomp}
\usepackage{wasysym}
\usepackage{setspace}
\usepackage{verbatim}
\usepackage{float}
\usepackage{ulem}
\usepackage[utf8]{inputenc}

\usepackage[text={17.2cm,25.2cm},centering]{geometry}
\numberwithin{equation}{section}

\begin{document}

\title{\textbf{Revisiting holographic model for thermal and dense QCD with a critical point}}

\author{Qingxuan Fu  $^{1}$\footnote{fuqx21@mails.jlu.edu.cn},
	Song He $^{1,2}$\footnote{hesong@jlu.edu.cn}, 
	Li Li $^{3,4,5}$\footnote{ liliphy@itp.ac.cn}, and
	Zhibin Li $^{6}$\footnote{lizhibin@zzu.edu.cn}}
\date{}
\maketitle
	
	\vspace{-10mm}
	
\begin{center}
	{\it
		$^{1}$ Center for Theoretical Physics and College of Physics, Jilin University,
		Changchun 130012, China\\ \vspace{1mm}			
		$^{2}$ Max Planck Institute for Gravitational Physics (Albert Einstein Institute), Am Muhlenberg 1, 14476 Golm, Germany\\ \vspace{1mm}			
		$^{3}$ CAS Key Laboratory of Theoretical Physics, Institute of Theoretical Physics,
		Chinese Academy of Sciences, Beijing 100190, China\\ \vspace{1mm}			
		$^{4}$ School of Fundamental Physics and Mathematical Sciences,
		Hangzhou Institute for Advanced Study, University of Chinese Academy of Sciences, Hangzhou 310024, China\\ \vspace{1mm}			
		$^{5}$ Peng Huanwu Collaborative Center for Research and Education, Beihang University, Beijing 100191, China\\ \vspace{1mm} 
		$^{6}$ Institute for Astrophysics, School of Physics, Zhengzhou University, Zhengzhou 450001, China\\ \vspace{1mm}
	}
	\vspace{10mm}
\end{center}

 \begin{abstract}
 To quantitatively provide reliable predictions for the hot and dense QCD matter, a holographic model should be adjusted to describe first-principles lattice results available at vanishing baryon chemical potential. The equation of state from two well-known lattice groups, the HotQCD collaboration and the Wuppertal-Budapest (WB) collaboration, shows visible differences at high temperatures. We revisit the Einstein-Maxwell-dilaton (EMD) holographic model for hot QCD with 2+1 flavors and physical quark masses by fitting lattice QCD data from the WB collaboration. Using the parameterization for the scalar potential and gauge coupling proposed in our work [Phys.Rev.D 106 (2022) 12, L121902], the equation of state, the higher order baryon number susceptibilities, and the chiral condensates are in quantitative agreement with state-of-the-art lattice results. We find that the critical endpoint (CEP) obtained from fitting the WB collaboration data is nearly identical to the one from the HotQCD collaboration, suggesting the robustness of the location of the CEP. Moreover, our holographic prediction for the CEP location is in accord with more recent Bayesian analysis on a large number of holographic EMD models and an effective potential approach of QCD from gap equations.
 \end{abstract}
	
	\baselineskip 18pt
	\thispagestyle{empty}
	\newpage
	
	\tableofcontents

\section{Introduction}\label{sec:01}

Studying the properties of the quark-gluon matter is essential to understanding the theory of the strong interactions known as Quantum Chromodynamics (QCD), which is the fundamental force describing the interactions between quarks and gluons. It is the main goal of heavy ion collision experiments at the Relativistic Heavy Ion Collider (RHIC) and the Large Hadron Collider (LHC). However, obtaining a quantitative understanding of the QCD phase structure at finite temperature $T$ and baryon chemical potential $\mu_B$ remains challenging due to the strongly coupled nature of the system under extreme conditions.

Several non-perturbative approaches have been proposed to study the QCD phase diagram under various conditions. Lattice QCD, formulated on a grid of points in space and time, provides a first-principle computation at zero baryon chemical potential and gives reliable information at small $\mu_B$ by extrapolating the lattice data. Lattice QCD suggests that the chiral and confinement/deconfinement phase transitions occur as an analytic crossover for small chemical potentials~\cite {Borsanyi:2010cj, Borsanyi:2013bia, HotQCD:2014kol}. On the other hand, effective theories such as the Dyson-Schwinger equation (DSE) ~\cite{Xin:2014ela, Gao:2016qkh,Qin:2010nq,Shi:2014zpa,Fischer:2014ata,Gao:2020qsj}, the Nambu-Jona-Lasinio (NJL) model~\cite{Asakawa:1989bq,Schwarz:1999dj,Li:2018ygx,Zhuang:2000ub} and the functional renormalization group (FRG)~\cite{Fu:2019hdw,Zhang:2017icm,Fu:2021oaw} have suggested that the crossover will become a first-order phase transition as $\mu_B$ increases. The critical point at the endpoint of the line of the first-order QCD phase transitions is known as the QCD critical endpoint (CEP). Despite decades of theoretical and experimental efforts, neither an exact location nor the properties of the CEP are well known. Nevertheless, currently lattice QCD results disfavor the existence of the CEP for $\mu_B/T\le 3$ and $\mu_B<300~\text{MeV}$~\cite{Vovchenko:2017gkg,Borsanyi:2020fev,Bazavov:2020bjn,Borsanyi:2021sxv,Bollweg:2022fqq,Philipsen:2021qji}.

An alternative non-perturbative approach is the so-called holographic QCD by the gauge gravity duality, which maps the strongly coupled non-Abelian gauge theories into a weakly coupled gravitational system with one higher dimension. It aims to capture the essential characteristics of realistic QCD and confront lattice QCD data at a quantitative level. Of particular interest is the Einstein-Maxwell-Dilaton (EMD) theory that was originally introduced in~\cite{DeWolfe:2010he, DeWolfe:2011ts}. The holographic EMD theory has been used to study various issues related to the quark matter in the literature, see~\cite{Chen:2022goa,Rougemont:2023gfz,Jarvinen:2022doa} for recent reviews. In recent studies, efforts have been made to improve the holographic QCD models and achieve a better quantitative description for hot QCD with 2+1 flavors and physical quark masses, see \emph{e.g.}~\cite{Knaute:2017opk,Critelli:2017oub,Cai:2022omk,Chen:2024ckb,Liu:2023pbt}. Non-perturbative effects and flavor dynamics are effectively incorporated into the model parameters by matching with lattice QCD data.
These improved models have shown qualitative consistency with the expected QCD phase diagram, including the existence of a CEP. Chiral condensation and gluon dynamics have also been studied, and the results are in quantitative agreement with lattice QCD simulations.

For holographic EMD models, the biggest difficulty is to find the scalar potential $V(\phi)$ and the coupling between scalar and gauge field $Z(\phi)$. As an effective field theory, the model parameters of the bulk gravitational theory should be fixed by matching with lattice QCD results, for which it is crucial to use up-to-date lattice simulation to make reliable predictions at finite $\mu_B$.  In our recent work~\cite{Cai:2022omk}, we provided a parameterization for $V(\phi)$ and $Z(\phi)$ with the five parameters fixed completely through lattice simulations at $\mu_B=0$: the equation of state and the second-order baryon number susceptibility. By adopting the lattice QCD data from the HotQCD collaboration for 2+1 quark flavors, we have obtained quantitative descriptions for QCD matter~\cite{Zhao:2022uxc,Li:2023mpv}. In particular, we predicted the exact location of the CEP with $(T_{CEP}, \mu_{CEP}) = (105, 555)~\text{MeV}$. A constraint on the holographic QCD phase transition from pulsar timing array observations can be found in~\cite{He:2023ado}.
Surprisingly, our parameterization is able to also capture hot and dense QCD matter for 2 flavors~\cite{Zhao:2023gur} and pure gluon~\cite{He:2022amv}. 

Besides the HotQCD collaboration, another mainstream lattice QCD group is the Wuppertal-Budapest collaboration (WB collaboration). The WB collaboration adopts a different setting compared to HotQCD, resulting in a different equation of state~\cite{Borsanyi:2018grb,Borsanyi:2013bia}. In particular, the equation of state at high $T$ starts to disagree even when considering the error bars. It is therefore natural to ask if the position of CEP is sensitive to the setting from both lattice QCD groups. In this study, we will use the parameterization in our previous work~\cite{Cai:2022omk} to fit the lattice data of WB collaboration at zero density~\cite{Borsanyi:2013bia}. We will compare our results for the higher order baryon number susceptibilities with the lattice computation from the WB collaboration~\cite{Borsanyi:2018grb}, as well as the equation of state at finite but small $\mu_B/T$~\cite{Borsanyi:2021sxv}. Both will give good quantitative agreement. Moreover, the position of the CEP obtained from fitting the WB collaboration data is nearly identical to the one we obtained by using the HotQCD data~\cite{Cai:2022omk}. It suggests that the location of CEP is pretty robust, not sensitive to the difference in the equation of state. Moreover, we will also show our holographic computation of the chiral condensation for the light quarks and the strange quark.

The subsequent sections of this work are organized as follows: 
In Section~\ref{sec:02}, we provide a brief introduction to our holographic setup and compare our holographic results to the lattice QCD data from the WB collaboration at zero baryon chemical potential.
Section~\ref{sec:muB} is dedicated to presenting the QCD phase structure derived from our holographic model, along with detailing the location of the CEP. In Section~\ref{sec:critical}, we delve into the computation of critical exponents in the vicinity of the CEP. We present our holographic computation of the chiral condensates for the light and strange quarks in Section~\ref{sec:chiral}. Finally, we provide a summary and discussion in Section~\ref{sec:discussion}. 

\section{Holographic setup and parameter fixing}\label{sec:02}

The gravitational action of the five-dimensional EMD theory is given by
\begin{align}
\begin{gathered}
S=\frac{1}{2 \kappa_N^2} \int d^5 x \sqrt{-g}\left[\mathcal{R}-\frac{1}{2} \nabla_\mu \phi \nabla^\mu \phi-\frac{Z(\phi)}{4} F_{\mu \nu} F^{\mu \nu}-V(\phi)\right]+S_\partial\,.
\end{gathered}
\label{eom}
\end{align}
where $\kappa_N^2$ is the effective Newton constant and $S_\partial$ is the boundary term. The dilaton $\phi$ is responsible for breaking the conformal invariance of boundary field theory and $A_\mu$ introduces the baryon chemical potential. Non-perturbative effects are effectively adopted into $V(\phi)$ and $Z(\phi)$ which are parameterized as~\cite{Cai:2022omk}
\begin{align}
& V(\phi)=-12 \cosh \left[c_1 \phi\right]+\left(6 c_1^2-\frac{3}{2}\right) \phi^2+c_2 \phi^6, \\
& Z(\phi)=\frac{1}{1+c_3} \operatorname{sech}\left[c_4 \phi^3\right]+\frac{c_3}{1+c_3} e^{-c_5 \phi}\,,
\end{align}
with five free parameters $c_1, c_2, c_3, c_4, c_5$.

The charged hairy planer black bole takes 
\begin{align}\label{ansatz}
\begin{gathered}
d s^2=-f(r) e^{-\eta(r)} d t^2+\frac{d r^2}{f(r)}+r^2 (d x^2+d y^2+d z^2)\,,\\
\phi=\phi(r),\quad A=A_t(r)dt\,,
\end{gathered}
\end{align}
with $r$ the holographic radial coordinate. The black hole event horizon is given by the largest root of $f(r_h)=0$, and the asymptotically AdS boundary is located at $r\rightarrow\infty$. The asymptotic expansion near the AdS boundary reads
\begin{equation}\label{UVdata}
\begin{split}
 \phi&=\frac{\phi_s}{r}+\cdots+\frac{\phi_v}{r^3}+\cdots,\quad  A_t=\mu_B-\frac{\kappa_N^2 n_B}{r^2}+\cdots\,,\\
 f&=r^2+\cdots+\frac{f_v}{r^2},\qquad\qquad\; \eta=0+\cdots\,,   
\end{split}
\end{equation}
with $\phi_s, \phi_v, \mu_B, n_B, f_v$ constants that will be fixed by solving the equations of motion. The source term $\phi_s$ essentially breaks the conformal symmetry and plays the role of the energy scale.

In our coordinate system, the boundary term for a well-defined Dirichlet variational principle and for removing divergence is given by
\begin{align}
\begin{aligned}
S_{\partial} & =S_{\text{Gibbons-Hawking}}+S_{\text {counter}} \\
& =\frac{1}{2 \kappa_N^2} \int_{r \rightarrow \infty} d x^4 \sqrt{-h}\left[2 K-6-\frac{1}{2} \phi^2-\frac{6 c_1^4-1}{12} \phi^4 \ln [r]-b \phi^4+\frac{1}{4} F_{\rho \lambda} F^{\rho \lambda} \ln [r]\right].
\end{aligned}
\end{align}
Here $h$ is the determinant of the induced metric at the AdS boundary and $K$ is the trace of the extrinsic curvature defined by the outward pointing normal vector to the boundary. The parameter $b$ is determined by requiring the pressure $P(T=0)=0$ at vanishing chemical potential. Via the standard holographic dictionary, we can then obtain many thermodynamic observables from the UV data~\eqref{UVdata}. In particular, $\mu_B$ and $n_B$ are baryon chemical potential and baryon number density, respectively. The free energy density, energy density, and pressure are given by
\begin{align}
\begin{aligned}
\Omega=&\frac{1}{2 \kappa_N^2}\left(f_v-\phi_s \phi_v-\frac{3-48 b-8 c_1^4}{48} \phi_s^4\right)\,, \\
\epsilon =&\frac{1}{2 \kappa_N^2}\left(-3 f_v+\phi_s \phi_v+\frac{1+48 b}{48} \phi_s^4\right)\,, \\
P =&\frac{1}{2 \kappa_N^2}\left(-f_v+\phi_s \phi_v+\frac{3-48 b-8 c_1^4}{48} \phi_s^4\right)\,.
\end{aligned}
\end{align}
Then one obtains the trace anomaly $I=\epsilon-3P$. Meanwhile, the temperature and entropy density can be obtained at the event horizon $r=r_h$.
\begin{align}
T=\frac{1}{4\pi}f'\left( r_h \right) e^{-\eta \left( r_h \right) /2}, \hspace{8mm} s=\frac{2 \pi}{\kappa_N^2} r_h^3\,.
\end{align}
The squared sound speed and the $n$-th order baryon number susceptibility are given by
\begin{align}\label{chincal}
c_s^2 =\left.\frac{d P}{d \epsilon}\right|_{\mu_B},  \hspace{8mm}
\chi_n^B(T,\mu_B)=\left. \frac{\partial^n}{\partial (\mu_B/T)^n}\frac{P}{T^4}\right|_T\,.
\end{align}
The baryon number susceptibilities are closely related to various cumulants of the baryon number distribution measured in heavy-ion collision experiments. In particular, at $\mu_B=0$ one has
\begin{align}
\chi_2^B(T,0)=\left. \frac{\partial^2}{\partial (\mu_B/T)^2}\frac{P}{T^4}\right|_T=\lim_{\mu_B\rightarrow 0} \frac{1}{T^2} \frac{n_B}{\mu_B}\,.
\end{align}
The hairy black hole solutions will be obtained numerically after further requiring the regularity at the event horizon. Please refer to~\cite{Cai:2022omk,Li:2020spf} for more details about the thermodynamics of hairy black holes.

\begin{figure}[htbp]
	\centering
	\includegraphics[width=0.45\textwidth]{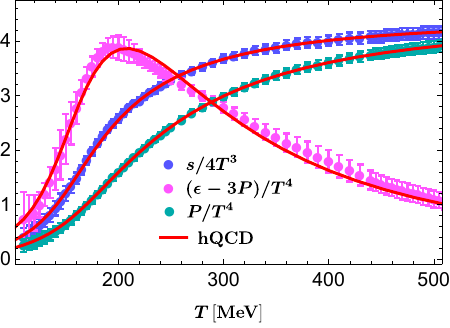}
	\qquad
	\centering
	\includegraphics[width=0.49\textwidth]{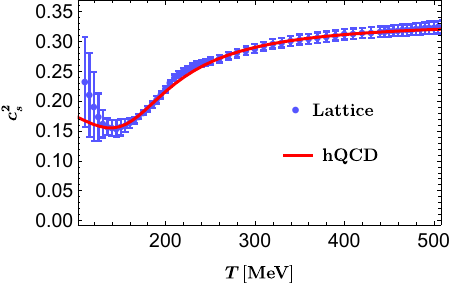}\\
	\caption{(Left) Dimensionless entropy density $s/T^3$, trace anomaly $(\epsilon-3P)/T^4$ and pressure $P/T^4$ with (Right) squared speed of sound $c_s^2$ at zero baryon density. Solid curves represent results from our holographic model, and data with error bars are lattice QCD data from WB collaboration \cite{Borsanyi:2013bia}.}
	\label{fig:0B1}
\end{figure}

As mentioned above, we fix all free parameters of the EMD model by matching the equation of state and the second-order baryon number susceptibility evaluated at $\mu_B=0$ with the corresponding lattice QCD data of WB collaboration~\cite{Borsanyi:2018grb,Borsanyi:2013bia}. We obtain that 
\begin{align}
\begin{aligned}
& c_1=0.7055,\quad c_2=0.0038,\quad c_3=1.7877,\quad c_4=0.1050,\quad c_5=30.2054\,, \\
& \phi_s=1080~\text{MeV},\quad \kappa_N{ }^2=(2 \pi) 1.76,\quad b=-0.2554\,.
\end{aligned}
\end{align}
These values are close to the one by fitting the lattice data of HotQCD collaboration~\cite{Cai:2022omk}, which is expected since the results from both lattice groups show small but visible differences at high temperatures.
The fitting results are depicted in Fig.~\ref{fig:0B1}. The left panel involves the comparison of the normalized entropy density, trace anomaly, and pressure obtained from our holographic model and the corresponding lattice data~\cite{Borsanyi:2013bia}. The right panel shows the squared speed of sound. One finds a good agreement with the lattice data over a considerable temperature range from $T=100~\text{MeV}$ to $500~\text{MeV}$. 

\begin{figure}[htbp]
	\centering
	\includegraphics[width=.48\textwidth]{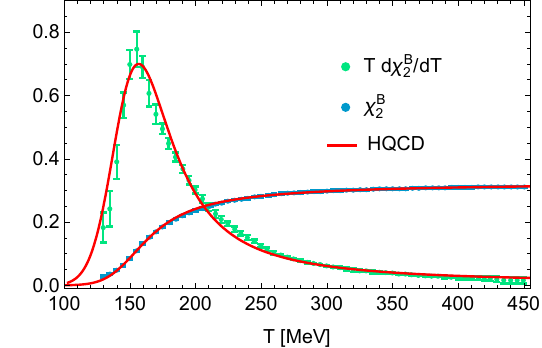}
	\quad
	\includegraphics[width=.48\textwidth]{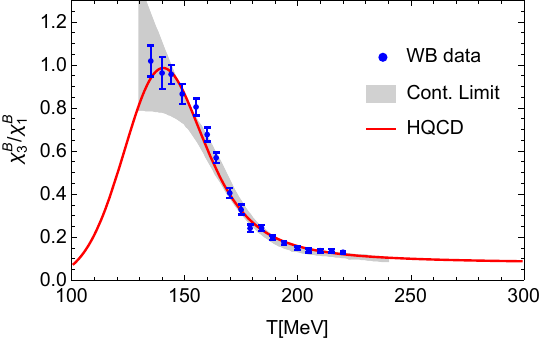}\\
	\includegraphics[width=.48\textwidth]{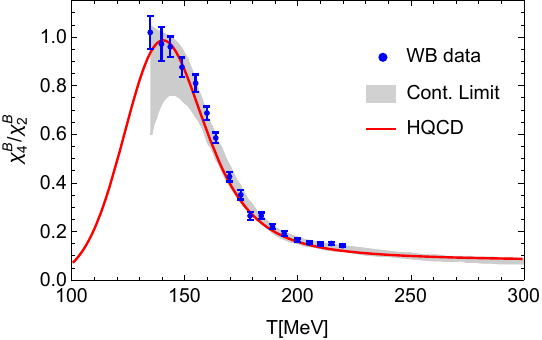}
	\;
	\includegraphics[width=.49\textwidth]{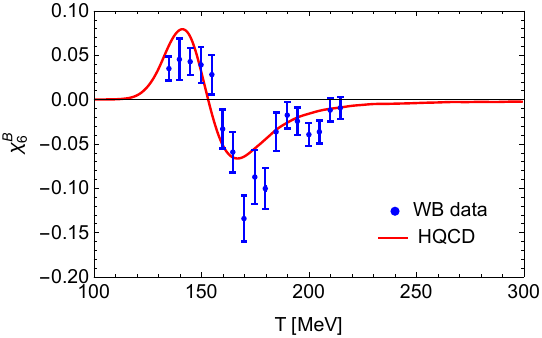}\\
	\caption{Baryon number susceptibilities $\chi^B_2$ and $T d\chi^B_2/dT$ (top left), $\chi^B_3/\chi^B_1$ (top right), $\chi^B_4/\chi^B_2$ (bottom left) and $\chi^B_6$ (bottom right) at $\mu_B=0$ compared with lattice data~\cite{Borsanyi:2013hza,Borsanyi:2018grb,Borsanyi:2021sxv}. The light-gray band denotes the region of continuous limit from lattice QCD simulation.}\label{figchin}
\end{figure}

Once obtaining the equation of state, one can compute the baryon number susceptibilities $\chi_n^B$ using~\eqref{chincal}. They are the coefficients in the Taylor expansion of $P$ around $\mu_B=0$, thus can reflect the information of the equation of state at small $\mu_B$. The baryon number susceptibilities at $\mu_B=0$ serve as an important probe to test the accuracy of our model. In Fig.~\ref{figchin}, we present a direct comparison between $\chi_n^B$ obtained from our holographic model and the corresponding lattice QCD computation from WB collaboration. The top left panel of Fig.~\ref{figchin} shows $\chi_2^B$ and $T d\chi_2^B/dT$, which we have used to determine the parameters $c_3, c_4, c_5$ in the coupling function $Z(\phi)$. The other three panels depict the temperature dependence of higher-order susceptibilities. In all cases, we observe good quantitative consistency, further demonstrating the accuracy of our holographic model.
%

\section{Thermodynamics and CEP at finite $\mu_B$}\label{sec:muB}

\begin{figure}[htbp]
	\centering
	\includegraphics[width=.455\textwidth]{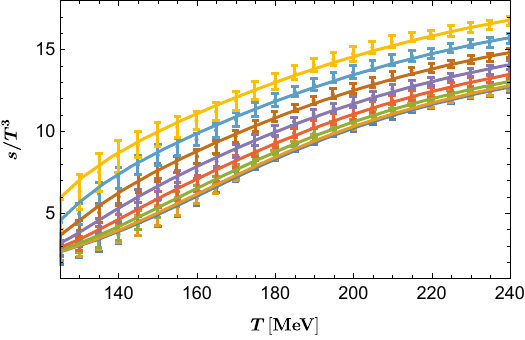}
	\qquad
	\includegraphics[width=.45\textwidth]{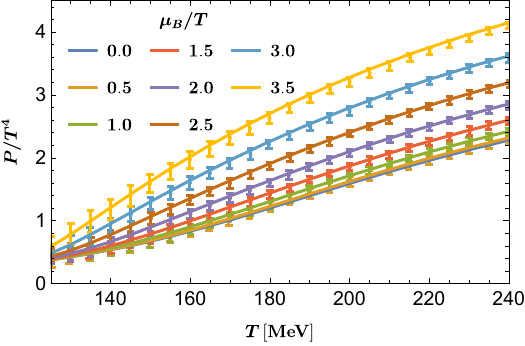}\\
	\hspace{-2mm}\includegraphics[width=.48\textwidth]{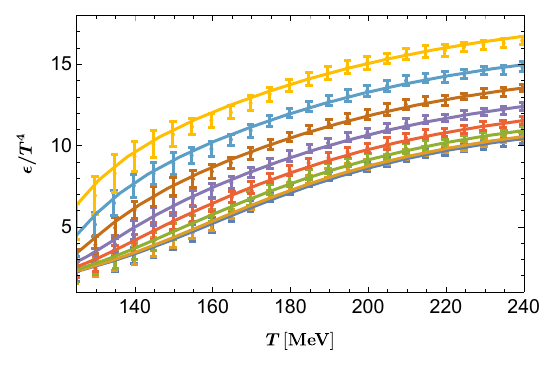}
	\qquad
	\hspace{-5mm}\includegraphics[width=.465\textwidth]{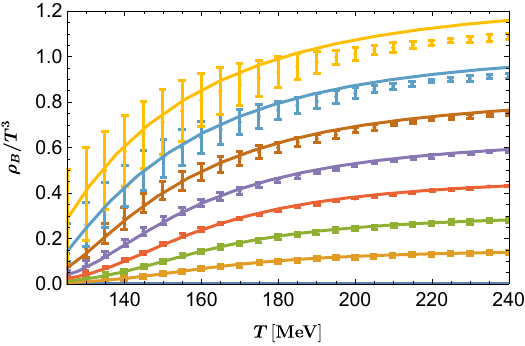}\\
	\caption{Dimensionless entropy density $s/T^3$, pressure $P/T^4$, baryon density $\rho/T^3$, and energy density $\epsilon/T^4$ as functions of temperature $T$ at various $\mu_B/T$. The solid curves obtained from our holographic model are compared with the lattice QCD data \cite{Borsanyi:2021sxv}, represented by dots with error bars.}
\label{fig:samallB}
\end{figure}

Using a novel expansion scheme, the most up-to-date lattice data for the QCD equation of state is available up to $\mu_B/T=3.5$~\cite{Borsanyi:2021sxv}. 
To further validate the accuracy of our holographic model, we compare the holographic prediction at finite $\mu_B$ with the one recently reported by~\cite{Borsanyi:2021sxv}. In Fig~\ref{fig:samallB}, we show the normalized entropy density, pressure, energy density, and baryon number density as a function of temperature for various $\mu_B/T$. Our holographic predictions for the entropy density, pressure, and energy density are in quantitative agreement with the lattice results all the way up to $\mu_B/T=3.5$. Regarding the baryon number density, there shows also quantitative agreement for most of the values of $T$ and $\mu_B$, although the holographic model overestimates the lattice results for $n_B$ at high temperatures $T\geq 195~\text{MeV}$ when $\mu_B/T\geq 3.0$. Nevertheless, this further supports the reliability and effectiveness of our holographic model in describing the thermodynamic properties of quark matter at nonzero $\mu_B$. 

\begin{figure}[htbp]
\centering
\includegraphics[width=.6\textwidth]{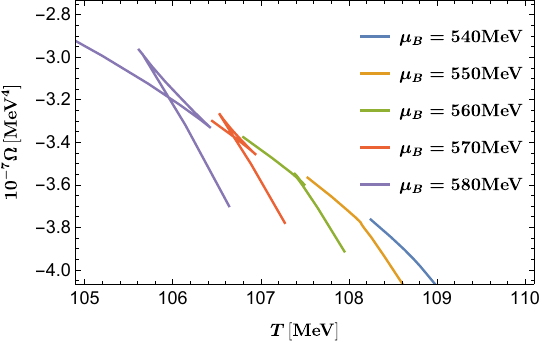}\\
\caption{Free energy density $\Omega$ as a function of temperature $T$ at different baryon chemical potential $\mu_B$. As $\mu_B$ increases, the smooth crossover becomes a first-order one.}
\label{fig:fre}
\end{figure}

After solving the bulk equations of
motion, we can obtain all extrema of the
free energy that corresponds to the coexistence region of not only thermodynamically stable minima but also metastable and unstable saddle points. The behavior of free energy density $\Omega$ versus temperature at fixed baryon chemical potential is depicted in Fig.~\ref{fig:fre}. At low $\mu_B$, $\Omega$ is a single-valued function of $T$, corresponding to a crossover without a clear phase boundary. In contrast, one observes
the characteristic multivalued swallow-shape for
$\Omega$, yielding a first-order phase transition. By identifying the physical state corresponding to the minimum free energy density, we can locate the boundary of the first-order phase transitions. 

\begin{figure}[htbp]
\centering
\includegraphics[width=0.75\textwidth]{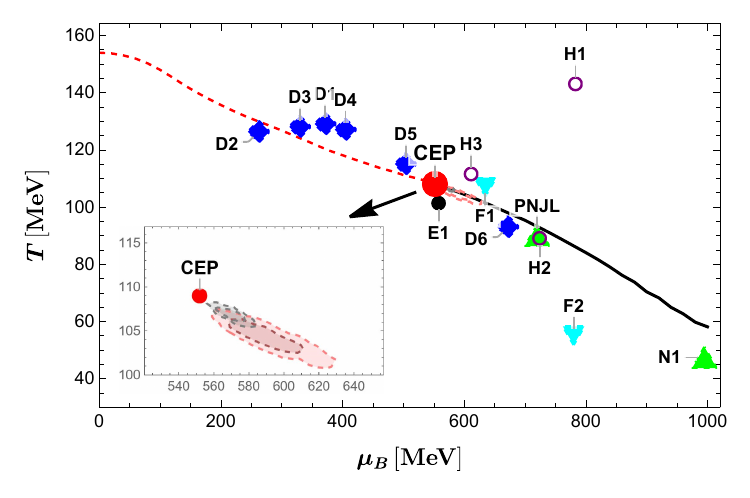}
\caption{Holographic prediction for the QCD phase diagram in $T$-$\mu_B$ plane. The red dashed line represents the crossover boundary, determined by the inflection point of $\chi_2^B$, and the black solid curve is the first-order transition line obtained by minimum free energy density. Between them there exists a critical endpoint(bold red dot) with $T_{CEP}=109~\text{MeV}$, $\mu_{CEP}=552~\text{MeV}$. CEP predicted by other QCD models is also presented, including D1-D6 from Dyson-Schwinger equation(DSE) \cite{Xin:2014ela,Gao:2016qkh,Qin:2010nq,Shi:2014zpa,Fischer:2014ata,Gao:2020qsj}, F1-F2 from functional renormalization group(FRG) \cite{Fu:2019hdw,Zhang:2017icm}, Polyakov-Nambu-Jona-Lasinio model \cite{Li:2018ygx}, N1 from Nambu-Jona-Lasinio \cite{Asakawa:1989bq}, E1 for effective potential approach~\cite{Zheng:2023tbv}, as well as H1-H3 from different holographic QCD models \cite{DeWolfe:2010he,Knaute:2017opk,Critelli:2017oub}. The pink and gray areas provide $68\%$ and $95\%$  confidence regions for the CEP location under different parametric ansatz \cite{Hippert:2023bel}. Surprisingly, the position of our CEP is completely within the $95\%$ confidence region.}
\label{fig:cep}
\end{figure}

The $T$-$\mu_B$ phase diagram obtained from our holographic approach is shown in Fig.~\ref{fig:cep}. The black solid line represents the line of the first-order phase transitions and the red dashed one denotes the inflection point of $\chi_2^B$ that characterizes the crossover region.
The CEP at the end of the first-order phase transition line is indicated by a red dot. More precisely, the location of the CEP is at $(T_{CEP}, \mu_{CEP}) = (109, 55 2)~\text{MeV}$. Remarkably, the CEP position obtained by fitting the WB lattice data is very close to the one we reported by fitting the hotQCD data, $(T_{CEP}, \mu_{CEP}) = (105, 555)~\text{MeV}$~\cite{Cai:2022omk}. In contrast, an early study~\cite{Critelli:2017oub} fitting the WB lattice data gave a significantly different location of the CEP, $(T_{CEP}, \mu_{CEP}) = (89, 724)~\text{MeV}$. Although a relatively complicated form of $V$ and $Z$ was adopted in~\cite{Critelli:2017oub} (see~\cite{Rougemont:2023gfz} for a review), our model improves the quality of the fitting for the same lattice data. For example, one can compare the trace anomaly in Fig.\,2 of~\cite{Grefa:2021qvt} with our case in the left panel of Fig.~\ref{fig:0B1} and the baryon number density in Fig.\,12 of~\cite{Grefa:2021qvt} with our case in Fig~\ref{fig:samallB}. We also note that the scaling dimension of the gauge theory operator dual to the dilaton field is $\Delta\approx 2.73$ of~\cite{Grefa:2021qvt}, while in our approach $\Delta=3$, which is much easier to extract the UV date from~\eqref{UVdata} numerically. A more recent study~\cite{Chen:2024ckb} incorporated the potential reconstruction method together with machine learning. The location of CEP for 2+1 flavors was found to be at $(T_{CEP}, \mu_{CEP}) = (94, 740)~\text{MeV}$ which is significantly from our results. However, as visible from Fig.\,4 of~\cite{Chen:2024ckb}, this holographic model overestimates the lattice results for the trace anomaly for $T\le 200\text{MeV}$. Meanwhile, the baryon number susceptibility $\chi^B_2$ displayed in Fig.\,5 of~\cite{Chen:2024ckb} yields a poor agreement with the lattice data. 

Nevertheless, it is worth noting that a recent study~\cite{Hippert:2023bel} shows that the predictions for the CEP location in different realizations of the model from a Bayesian analysis overlap at one sigma, with $T_{CEP} = 101-108\ \text{MeV}$ and $\mu_{CEP} = 560-625\ \text{MeV}$.\footnote{Note that the location of the CEP reported by~\cite{Chen:2024ckb} is outside the one sigma region of~\cite{Hippert:2023bel}. A better choice of function configurations for the deformed factor and gauge coupling function in~\cite{Chen:2024ckb} could give potential improvement on the fitting quality with lattice QCD.} Meanwhile, the CEP from an effective potential approach is found to be located at $(T_{CEP}, \mu_{CEP}) = (101.3, 558)~\text{MeV}$~\cite{Zheng:2023tbv}. Both agree with our CEP, suggesting that our prediction for the location of the CEP is robust. 

\section{Critical phenomena near the CEP}\label{sec:critical}

The critical exponents are vital parameters used to describe critical points and are regarded as universal physical quantities. They transcend the specifics of a physical system and depend solely on the system's degrees of freedom and correlation length. As one approaches the critical endpoint (CEP), thermodynamic quantities often exhibit power-law scaling with temperature or chemical potential. Consequently, along different directions, a set of critical exponents can be defined to provide a more nuanced characterization of the system's critical behavior.

\begin{figure}[htbp]
	\centering
	\includegraphics[width=.46\textwidth]{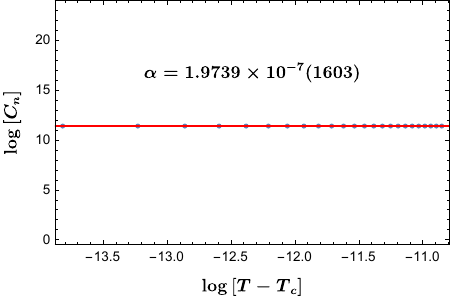}
	\qquad
	\includegraphics[width=.465\textwidth]{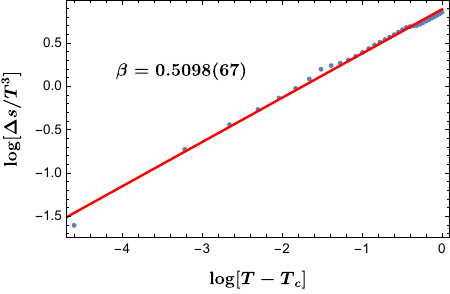}\\
	\includegraphics[width=.45\textwidth]{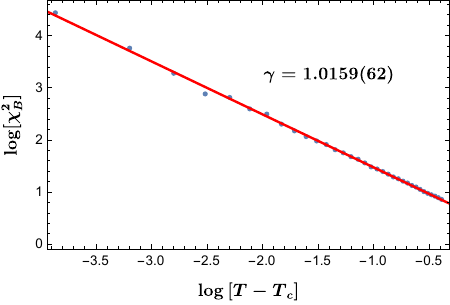}
	\qquad
	\includegraphics[width=.465\textwidth]{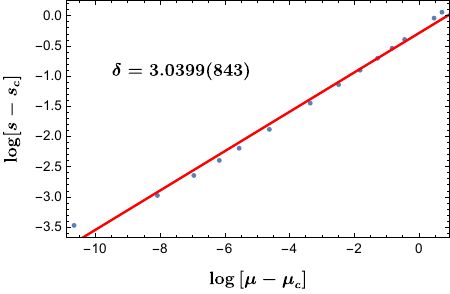}\\
	\caption{The critical exponents $\alpha$, $\beta$, $\gamma$ and $\delta$ from our holographic model.}
	\label{fig:exponents}
\end{figure}

The exponent $\alpha$ and $\gamma$ are defined along the first-order axis which is the tangent of the first-order line near the CEP, with the power law of the specific heat and baryon number susceptibility, respectively. 
\begin{equation}\label{alpha}
    C_{\rho}\equiv T\left(\frac{\partial s}{\partial T}\right)_{\rho_B}=-T\left( \frac{\partial^2\Omega}{\partial T^2}-\frac{(\partial^2\Omega/\partial T\partial\mu)^2}{(\partial^2 \Omega/\partial \mu^2)}\right)\sim|T-T_{\text{CEP}}|^{-\alpha}.
\end{equation}
\begin{equation}\label{gamma}
    \chi^B_2=\frac{1}{T^2}\left( \frac{\partial n_B}{\partial \mu_B}\right)_T\sim |T-T_{\text{CEP}}|^{-\gamma}.
\end{equation}
On either side of the first-order phase transition line, which is determined by the minimum free energy, thermodynamic quantities display two distinct branches corresponding to the high-temperature and low-temperature phases. These branches typically exhibit discontinuities at the phase boundary, enabling us to observe significant changes in various thermodynamic quantities during the first-order phase transition. The critical exponent $\beta$ is related to the power-law relationship of the discontinuous change in entropy density $s$ along the first-order phase transition line and the temperature $T$.
\begin{equation}\label{beta}
    \Delta s=s_>-s_<\sim (T_{\text{CEP}}-T)^\beta.
\end{equation}
Where $s_>$ and $s_<$ represent entropy density from the high-temperature and low-temperature branches, respectively. And the critical exponent $\delta$ is defined follow the relation of $s-s_{\text{CEP}}$ and $\mu_B-\mu_{\text{CEP}}$ at the critical isotherm $T=T_{\text{CEP}}$.
\begin{equation}
    s-s_{\text{CEP}}\sim|\mu_B-\mu_{\text{CEP}}|^{1/\delta}\,.
\end{equation}
Fig.~\ref{fig:exponents} shows the behavior of different thermodynamic quantities near the critical endpoint. And critical exponents are obtained by fitting the slope of different log plots. 
The values of the four critical exponents are shown in Table~\ref{tabexp}. We also compared the values of the critical exponents near CEP given by our holographic model and the non-QCD fluids experiment data, the results of 3D Ising model, the mean-field theory, and the holographic QCD model (the DGR model)~\cite{Goldenfeld:1992qy, DeWolfe:2010he}. Surprisingly, our results are closest to the results of the mean-field theory, which is consistent with the results of the holographic QCD model~\cite{DeWolfe:2010he}.
\begin{table}[ht!]
    \centering
    \setlength{\tabcolsep}{3.5mm}{
    \begin{tabular}{|c|c|c|c|c|c|}
     \cline{1-6}
      &  Experiment      &  3D Ising         & Mean field & DGR model & HQCD   \\ \cline{1-6}
    $\alpha$ & 0.110-0.116 &  0.110(5)         &  0         &   0          & 1.9739 $\times$ $10^{-7}$(1603) \\ \cline{1-6}
    $\beta$  & 0.316-0.327 &  0.325$\pm$0.0015 & 1/2        & 0.482        & 0.5098(67) \\ \cline{1-6}
    $\gamma$ &  1.23-1.25  &  1.2405$\pm$0.0015& 1          & 0.942        & 1.0159(62)  \\ \cline{1-6} 
    $\delta$ &  4.6-4.9    &  4.82(4)          & 3          & 3.035     &    3.0399(843)\\ \cline{1-6} 
    \end{tabular}}
    \caption{Critical exponents from experiments in non-QCD fluids, the full quantum 3D Ising model, mean-field (van der Waals) theory, the DGR model, and our 2+1-flavor holographic model.}
    \label{tabexp}
\end{table}
%

\section{Chiral condensate}\label{sec:chiral}

To incorporate flavor dynamics into our model, we adopt techniques akin to those in the KKSS model~\cite{Karch:2006pv}. This involves introducing a holographic probe action to describe the chiral condensates. With the hairy bulk geometry~\eqref{ansatz} for the (2+1)-flavor system, we adopt the following effective action for the light quakes ($u$ and $d$ quarks) and the strange quark ($s$ quark).
\begin{align}
\begin{gathered}
S_{X_q}=\frac{1}{2 \kappa_N^2} \int d^5 x \sqrt{-g} Z_q(\phi)\left[-\frac{1}{2} \nabla_\mu X_q \nabla^\mu X_q-V\left(X_q\right)\right]+S_{X_q, \partial}\,, 
\end{gathered}
\end{align}
with $q=l$ for the light quarks and $q=s$ for the $s$ quark. The boundary term $S_{X_q, \partial}$ is essential to obtain the physical condensation. The bulk scalar field $X_q$ is dual to the chiral operator $\bar{\psi}_q \psi_q$ with the scaling dimension $\Delta_c=3$.
Following our previous work~\cite{Cai:2022omk}, the coupling functions $Z_q(\phi)$ and $V(X_q)$ take 
\begin{align}
\begin{gathered}
Z_q(\phi)=a_1 e^{a_q \phi^2}, \quad V\left(X_q\right)=-\frac{3}{2} X_q^2+a_0 X_q^4,\quad q=l, s\,,
\end{gathered}
\end{align}
where $a_0, a_1, a_q$ are parameters that will be fixed by fitting lattice QCD data. 

\begin{figure}[htbp]
\centering
\includegraphics[width=0.48\textwidth]{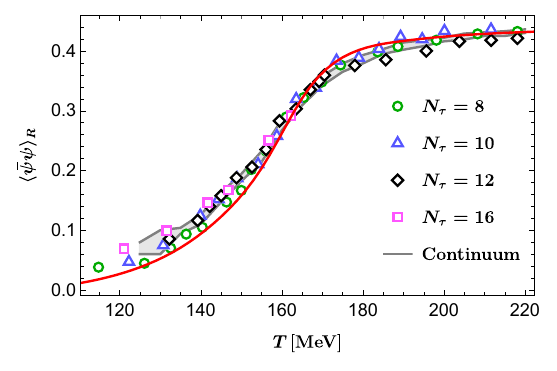}\;
\includegraphics[width=0.48\textwidth]{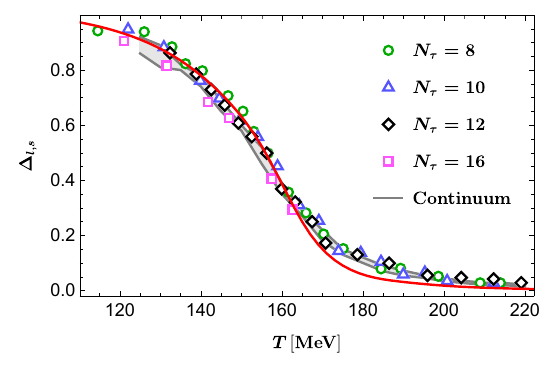}
\caption{Renormalized chiral condensate and subtracted chiral condensate compare with lattice data \cite{Borsanyi:2010bp}.}
\label{fig:Deltals}
\end{figure}

The equation of motion on top of~~\eqref{ansatz} reads,
\begin{align}
X_q^{\prime \prime}+\left(\frac{f^{\prime}}{f}-\frac{\eta^{\prime}}{2}+\frac{3}{r}\right) X_q^{\prime}+\frac{\partial_\phi Z_q X_q{ }^{\prime} \phi^{\prime}}{Z_q}-\frac{1}{f} \partial_{X_q} V=0\,,
\label{chiralS}
\end{align}
and the behavior of $X_q$ near the AdS boundary is given by
\begin{align}
X_q(r)=\frac{m_q}{r}+\cdots+\frac{\sigma_q}{r^3}+\cdots.
\end{align}
Here $\sigma_q$ is a constant and $m_q$ is nothing but the mass of each quark. Matching lattice QCD data, we choose $m_l=5.1 \mathrm{MeV}$ and $m_s=102 \mathrm{MeV}$. The chiral action suffers from divergence and should be renormalized with a counter term that takes
\begin{align}
\begin{aligned}
S_{X_q, \partial}=\frac{1}{2 \kappa_N^2} \int_{r \rightarrow \infty} d x^4 \sqrt{-h}\left[-\frac{1}{2} a_1 X_q^2+a_1 a_0 X_q^4 \ln[r]+\frac{a_1\left(1-6 a_q\right)}{6} X_q^2 \phi^2 \ln [r]\right]\,.
\end{aligned}
\end{align}
Then we can obtain the chiral condensate
\begin{align}
\begin{aligned}
\langle\bar{\psi} \psi\rangle_{q, T}  =\frac{\delta\left(S_{X_q}+S_{X_q, \partial}\right)}{\delta m_q}  =\frac{a_1}{2 \kappa_N^2}\left[2 \sigma_q+2 a_0 m_q^3+\frac{1}{4} m_q \phi_s^2\right] .
\end{aligned}
\end{align}
To compare directly with the W-B lattice simulation, we define the renormalized chiral condensate and subtracted chiral condensate as \cite{Borsanyi:2010bp}
\begin{align}
\begin{aligned}
\langle\bar{\psi} \psi\rangle_{R} =-\frac{m_l}{X^4}\left[ \langle\bar{\psi} \psi\rangle_{l, T} -\langle\bar{\psi} \psi\rangle_{l, 0} \right]~\text{and}~~
\Delta_{l,s}=\frac{ \langle\bar{\psi} \psi\rangle_{l, T}-\frac{m_l}{m_s}\langle\bar{\psi} \psi\rangle_{s, T}}{\langle\bar{\psi} \psi\rangle_{l, 0}-\frac{m_l}{m_s}\langle\bar{\psi} \psi\rangle_{s, 0}}.
\end{aligned}
\end{align}
with $\langle\bar{\psi} \psi\rangle_{q, 0}$ the chairl condensate at $T=0$.\footnote{$\langle\bar{\psi} \psi\rangle_{q, 0}$ can be obtained either by treated as a free parameter to be adjusted to fit lattice data or by evaluating $\langle\bar{\psi} \psi\rangle_{q, T}$ at sufficiently low temperatures. Both yield identical results, $\langle\bar{\psi} \psi\rangle_{l, 0}=1.81 \times 10^7$, $\langle\bar{\psi} \psi\rangle_{s, 0}=-1.033 \times 10^8$.} The temperature dependence of the renormalized chiral condensate and subtracted chiral condensate are depicted in Fig.~\ref{fig:Deltals} with $a_0=13, a_1=2.76, a_l=0.58, a_s=0.8$ and $X=m_{\pi}=135$ MeV. One finds a remarkably quantitative agreement with lattice data~\cite{Borsanyi:2010bp}. It could serve as a strong support for our holographic setup.

\begin{figure}[htbp]
	\centering
	\includegraphics[width=0.47\textwidth]{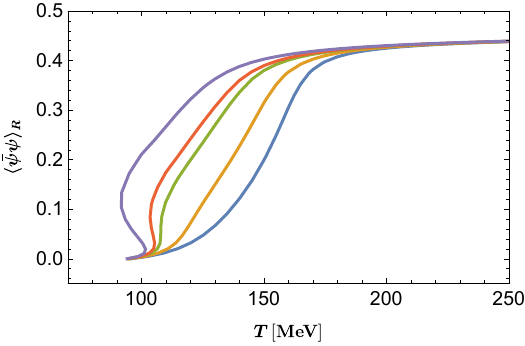}
	\qquad
	\includegraphics[width=0.47\textwidth]{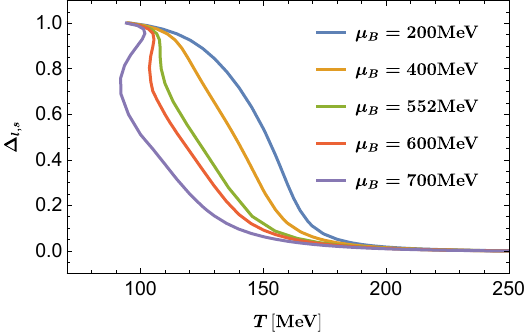}
	\vspace{-5mm}\caption{Temperature dependence of the renormalized chiral condensate and subtracted chiral condensate for different values of $\mu_B$. The characteristic multivalued S-shape develops when $\mu_B$ is beyond the critical value $\mu_{CEP}=552~\text{MeV}$.}
	\label{fig:lsfb}
\end{figure}

{Our holographic predictions for the renormalized chiral condensate $\langle\bar{\psi} \psi\rangle_{R}$ and subtracted chiral condensate $\Delta_{l,s}$ as a function of temperature are displayed in Fig.~\ref{fig:lsfb} along lines of constant $\mu_B$. One finds that both the renormalized chiral condensate $\langle\bar{\psi} \psi\rangle_{R}$ and subtracted chiral condensate $\Delta_{l,s}$ for different $\mu_B$ tend to a fixed value at high temperatures. $\langle\bar{\psi} \psi\rangle_{R}$ is promoted while $\Delta_{l,s}$ is suppressed as the baryon chemical potential is increased. For a given $\mu_B$, $\Delta_{l,s}$ is larger at low temperatures and decreases as the temperature increases. At high temperatures, the renormalized chiral condensates $\langle\bar{\psi} \psi\rangle_{R}$ all decrease to negative values. One also notices the formation of a characteristic multivalued S-shape for both chiral condensates when $\mu_B$ is larger than the critical value $\mu_{CEP}=552~\text{MeV}$, suggesting the development of a first-order phase transition.}
%

\section{Conclusion and discussion}\label{sec:discussion}

The motivation for our study arises from a unique situation in lattice QCD: there exist only two sets of lattice configurations that provide the equation of state (EoS) for QCD matter at zero chemical potential and finite temperature. While these lattice QCD results are quantitatively consistent in the low-temperature regime, they exhibit visible discrepancies at high temperatures. The discrepancies stem from differences in the choice of fermion propagators and initial configurations employed by the two lattice groups, resulting in markedly distinct EoS for hot QCD matter.

Currently, there is no definitive criterion to discern the discrepancies between these two sets of lattice data. To address this issue, we have previously developed a holographic QCD model that successfully reproduces the EoS consistent with one of these lattice groups, known as "hot-QCD." Given the inconsistent behavior of lattice data in the high-temperature region, it becomes imperative to establish another holographic QCD model that quantitatively predicts the critical point in the QCD phase diagram based on the data from the other lattice group, referred to as "WB".

The precise location of this critical point is of paramount importance, as it exhibits high sensitivity to the model parameters. Thus, the primary motivation behind this work is to thoroughly investigate and determine the exact position of the critical point, addressing the discrepancies in the lattice data and advancing our understanding of the QCD phase diagram.

In this work, we build up a bottom-up holographic model to confront the most recent lattice results for EOS from the different lattice QCD collaboration and offer a reliable first-order transition line and CEP in the QCD phase diagram. By computing thermodynamic quantities via holographic renormalization, the EOS is found to be quantitatively matched with the latest lattice QCD simulation. The QCD first-order transition line is fixed from the free energy and the corresponding CEP in the $T-\mu_B$ plane is predicted at $(T_C=109 \text{MeV}, \mu_C=552 \text{MeV})$.
The CEP in the 2+1 flavor holographic QCD model, calibrated using data from the Wuppertal-Budapest lattice QCD group, aligns closely with predictions from a separate holographic model established by the HotQCD collaboration \cite{Cai:2022omk}. This proximity is also mirrored in the Bayesian analysis results on the location of the QCD critical point from a holographic perspective \cite{Hippert:2023bel}, indicating that the location of the CEP exhibits low sensitivity to variations in lattice QCD data inputs. Additionally, the associated critical exponents approximate those expected from mean-field theory. Contrarily, as demonstrated by \cite{Zhao:2023gur}, the holographic 2-flavor CEP diverges significantly from mean-field predictions, underscoring the distinctiveness of our holographic approach compared to conventional large N QCD models. Our development of diverse holographic models, tailored to mimic specific QCD systems (e.g., 2+1 flavor\cite{Cai:2022omk}, pure gluon \cite{He:2022amv} and 2-flavor \cite{Zhao:2023gur}) based on corresponding lattice QCD data, enables precise microscopic descriptions. The similarity in critical exponents suggests that the 2+1-flavor model may belong to the same critical class as the mean-field theory. Finally, we investigate the behavior of renormalized chiral condensate and subtracted chiral condensate which is consistent with the WB lattice QCD simulation \cite{Borsanyi:2010bp}. 

Dedicating to a precise characterization of the properties and differences of the two phases along the first-order phase transition is an interesting direction for further study. In the current investigation, we have set up the preliminary hQCD model to quantitatively study the phase transition in the QCD phase diagram. The current model captures the characteristic confining properties and many other relevant physical quantities should be taken into account to complete the phase diagram, including the chiral condensation and the transport properties in QGP and hadron gas. In addition, the present results, in particular those regarding the CEP, should be embedded into the framework of a more general and multidimensional view of the QCD phase diagram, including an external magnetic field, an isospin chemical potential, and a rotation. It will be interesting to consider the real-time dynamics in our hQCD model far from equilibrium.

\section{Acknowledgments}\label{sec:06}

This work is supported by the NSFC Grants No.12075101 and No.12235016, No.12122513, No.12075298, and No.12047503. S.H. would also appreciate financial support from the Fundamental Research Funds for the Central Universities and Max Planck Partner Group. Z.L. acknowledges support from the National Key Program for Science and Technology Research Development (2023YFB3002500) and the Natural Science Foundation of Henan Province of China Grant No. 242300420235.


\providecommand{\href}[2]{#2}\begingroup\raggedright\endgroup
\end{document}